\documentclass[aps,prd,showpacs,floatfix,twocolumn,nofootinbib,superscriptaddress]{revtex4-1} %
\usepackage{dcolumn}% Align table columns on decimal point
\usepackage{amsmath,amsfonts,amssymb,mathrsfs,times,bm}
\usepackage{extarrows}
\usepackage{graphicx}
\usepackage{float}
\usepackage{booktabs}
\usepackage{graphicx,epstopdf}% Include figure files
\usepackage{dcolumn}% Align table columns on decimal point
\usepackage{exscale}
\usepackage{relsize}
\usepackage{latexsym}
\usepackage{longtable}
\usepackage[colorlinks=true,breaklinks=true,linkcolor=blue,citecolor=blue,urlcolor=blue]{hyperref}

\newcommand{\nn}{\nonumber}

\newcommand{\orcid}[1]{\href{https://orcid.org/#1}{\includegraphics[scale=0.035]{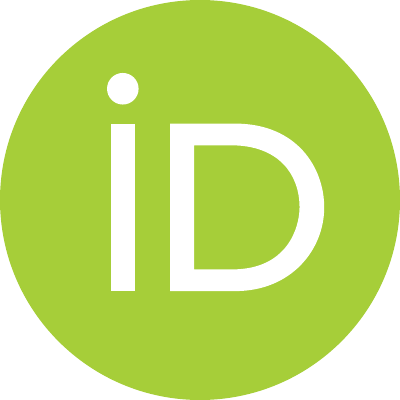}}}

\begin{document}

\title{Leading-order deflection of particles by a moving Schwarzschild lens with \\a two-dimensional velocity}

\author{Xuan Wang}
\affiliation{School of Mathematics and Physics, University of South China, Hengyang 421001, China}
\author{Wenbin Lin\hspace*{0.6pt}\orcid{0000-0002-4282-066X}\hspace*{0.8pt}}
\affiliation{School of Mathematics and Physics, University of South China, Hengyang 421001, China}
\affiliation{School of Physical Science and Technology, Southwest Jiaotong University, Chengdu 610031, China}
\author{Ghulam Mustafa\hspace*{0.6pt}\orcid{0000-0003-1409-2009}\hspace*{0.8pt}}
\affiliation{Department of Physics, Zhejiang Normal University, Jinhua 321004, China}
\author{Guansheng He\hspace*{0.6pt}\orcid{0000-0002-6145-0449}\hspace*{0.8pt}}
\email{Corresponding author. hgs@usc.edu.cn}
\affiliation{School of Mathematics and Physics, University of South China, Hengyang 421001, China}
\affiliation{Purple Mountain Observatory, Chinese Academy of Sciences, Nanjing 210023, China}

\date{\today}

\begin{abstract}
The gravitational deflection effect of relativistic massive and massless particles up to the first post-Minkowskian order caused by a moving Schwarzschild black hole with a two-dimensional equatorial velocity, which contains the radial and transversal components, is studied analytically, and a new unified formula for the deflection angle is achieved. The expression of the angle matches well with the results of the weak deflection of relativistic particles induced by a radially moving Schwarzschild source given in the literature, when the transversal component of the lens velocity vanishes. The joint velocity effect, which consists of the influences of the transversal and radial motions of the lens on the leading-order Schwarzschild deflection of the massive particles and light, is then discussed in the context of general relativity. We analyze the order of magnitude of this kinematical effect and evaluate the possibility of its astronomical detection subsequently.

\begin{description}

\item[Keywords]
Velocity effects; Gravitational lens; Black holes
\end{description}

\end{abstract}

\maketitle

\section{Introduction}  \label{sect1}
Due to its fruitful applications in astronomy, gravitational lensing (GL) has developed into one of the most rapidly growing branches of modern astrophysics~\cite{BN1992,Wambsganss1998,VE2000,KP2005,BSN2007,Virbha2009,TKNA2014,ZX2016,ZX2017,FLBPZ2017,JOSV2018,JL2019,LM2019,CGV2019,JBO2019,GX2021,LLJ2021,HXJL2024,JLRV2024}. In the last decades, considerable attention from the relativity community has been paid to the study of the GL phenomena induced by a moving gravitational system (see, for instance,~\cite{BG1983,PB1993,KS1999,KP2003,WS2004,Heyrovsk2005,KM2007,KF2007,DX2012,BZ2015,Deng2015,Zscho2018,Zscho2019} and references therein). The reasons responsible for an increasing interest on this kind of GL effects lie mainly in two aspects. Firstly, we know that the key to study this type of GL effects depends on the knowledge of the influence of the lens motion relative to the observer on propagation of test particles and on the related relativistic effects, which is the so-called velocity effect~\cite{PB1993,WS2004,Heyrovsk2005}. When a gravitational lens moves with a large (or even relativistic) velocity, the velocity effects will become so evident that they may affect the high-accuracy measurements of the observable lensing properties of the images. It is thus necessary to further probe the velocity effects on propagation of test particles and on the lensing observables for various astronomical scenarios and different background geometries. On the other aspect, great progress in techniques and instruments of high-accuracy astronomical observation (see, e.g.,~\cite{Perryman2001,SN2009,Reid2009,Trippe2010,Malbet2012,ZRMZBDX2013,RH2014,Malbet2014,Prusti2016,Murphy2018,RD2020,Brown2021,LXLWBLYHL2022,LXBLLLH2022}) has been made over the last decades. The astrometric precision in current surveys and forthcoming new telescope networks for multiwavelength observations is at the level of $1\!\sim\!10$ microarcseconds ($\mu$as) or better~\cite{SN2009,Reid2009,Malbet2014,Murphy2018}. For example, the Square Kilometre Array (SKA)~\cite{BBGKW2015,LXLWBLYHL2022} and other next-generation radio observatories (see, e.g.,~\cite{Murphy2018,RD2020}) aim at an angular accuracy of about $1\mu$as. The planned Nearby Earth Astrometric Telescope (NEAT) mission~\cite{Malbet2012,Malbet2014} is working towards an unprecedented space-borne astrometric accuracy of $0.05\,\mu$as. Additionally, the angular resolution of current astronomical detectors for observing massive particles or multi-messengers can reach one degree or better~\cite{Aab2014,Albert2020}. For instance, the angular resolution of the ANTARES neutrino telescope is about $0.59^{\circ}\pm0.10^{\circ}$ for downward-going muons~\cite{Albert2020}. With the continual improvement in astronomical measurements and the increase in the multi-messenger synergic observations~\cite{BT2017,MFHM2019,IceCube2018,QJFZZ2021}, a further consideration of the velocity effects on the GL phenomena of various photonic or nonphotonic messengers becomes more and more important.

To the best of our knowledge, most of efforts were made to discuss the velocity effects on the propagation and gravitational lensing of light signals in the weak-field limit, and only few works were devoted to considering the motion effect of the gravitational source on the lensing properties of timelike signals~\cite{WS2004,HL2014,HL2016a,HL2017b}. In 2004, Wucknitz and Sperhake~\cite{WS2004} performed a detailed discussion of the radial motion effect of a moving Schwarzschild lens on the gravitational deflection of particles and photons up to the first post-Minkowskian (PM) order. The gravitational deflection of relativistic massive particles and light propagating in the equatorial plane of a radially moving Kerr-Newman black hole, along with the effects of the radial lens motion on the second-order contributions to the deflection angle, was also investigated in~\cite{HL2016a,HL2017b}. However, a full theoretical treatment of the joint velocity effects induced by a more general motion of the lens on the gravitational deflection of test particles including photons has not been performed up to now.

In present work, we adopt the iterative technique in~\cite{WS2004,HL2017b} and compute the leading-order gravitational deflection of relativistic massive particles and light induced by a moving Schwarzschild lens with a two-dimensional equatorial constant velocity in the 1PM approximation. The resulting bending angle, which takes a unified and compact form, extends the results given in the literature~\cite{WS2004,HL2017b}, since the full effect of the transversal motion of the lens, along with its radial motion effect, is also included into the kinematical correctional coefficient of the angle. We then perform a fully general-relativistic discussion of the joint transversal and radial velocity effect on the first-order deflection of the massive and massless particles in Schwarzschild spacetime, where the order of magnitude of this kinematical effect and the possibility to detect it are estimated. Our discussions are limited in the weak-field, small-angle, and thin-lens approximation.

This paper is organized as follows. We review the Kerr-Schild form of the weak-field metric of a moving Schwarzschild black hole with a two-dimensional constant velocity in Sec.~\ref{sect2}, and derive the equations of motion of test particles traveling in the equatorial plane of the moving Schwarzschild source within the 1PM approximation in Sec.~\ref{sect3}. Section~\ref{sect4} is devoted to the calculation of the leading-order gravitational deflection of relativistic particles including photons due to the moving lens, and the transversal and radial velocity effects on the first-order Schwarzschild deflection of the particles are probed in Sec.~\ref{sect5}. A summary is given in Section~\ref{sect6}. Throughout this work, the geometrized units in which $G=c=1$ are adopted, Greek indices run over $0,~1,~2$, and $3$, and Latin indices run over $1,~2$, and $3$.

\section{The metric of a moving Schwarzschild black hole with a two-dimensional velocity} \label{sect2}

Let $\bm{e}_i \left(i=1,~2,~3\right)$ be the orthonormal basis of a three-dimensional Cartesian coordinate system. The rest Kerr-Schild coordinate frame of the observer of the background and the comoving Kerr-Schild coordinate frame of the gravitational source are denoted by $x_\nu=\left(t,~x,~y,~z\right)$ and $X_\nu=(T,~X,~Y,~Z)$, respectively. Then the metric of a moving Schwarzschild black hole with an arbitrary two-dimensional constant velocity $\bm{v}=v_1\bm{e}_1+v_2\bm{e}_2$ within the 1PM approximation in the observer's rest Kerr-Schild frame $\left(t,~x,~y,~z\right)$ reads~\cite{LFZMH2021}:
\begin{widetext}
\begin{eqnarray}
&& g_{00}=-1+\frac{2\gamma^2M\left(R-v_1X-v_2Y\right)^2}{R^3}+\mathcal{O}\!\left(M^2\right)~,  \label{g00} \\
&& g_{0i}=\frac{2\gamma M\left(R-v_1X-v_2Y\right)}{R^2}\left\{\!\frac{X_i}{R}\!-\!v_i\!\left[\gamma\!-\!\frac{\left(\gamma\!-\!1\right)\left(v_1X\!+\!v_2Y\right)}{v^2R}\right]\!\right\}
+\mathcal{O}\!\left(M^2\right)~,~~~~ \label{g0i}  \\
&&g_{ij}=\delta_{ij}+\frac{2M}{R}\!\left\{\frac{X_i}{R}-v_i\!\left[\gamma-\frac{\left(\gamma-1\right)\left(v_1X+v_2Y\right)}{v^2R}\right]\right\}
\left\{\!\frac{X_j}{R}\!-\!v_j\!\left[\gamma\!-\!\frac{\left(\gamma\!-\!1\right)\left(v_1X\!+\!v_2Y\right)}{v^2R}\right]\!\right\}+\mathcal{O}\!\left(M^2\right)~,~~~~  \label{gij}
\end{eqnarray}
\end{widetext}
where $M$ denotes the rest mass of the black hole, $R=\sqrt{X^2+Y^2+Z^2}$, $v=\sqrt{v_1^2+v_2^2}$, $\gamma\equiv\left(1-v^2\right)^{-\scriptstyle\frac{1}{2}}$ is the Lorentz factor, $\delta_{ij}$ stands for the Kronecker delta, and the coordinates of the observer's rest Kerr-Schild frame are related to the ones of the comoving Kerr-Schild frame by the following Lorentz transformation:
\begin{eqnarray}
&& T=\gamma\left(t-v_1x-v_2y\right)~,  \\
&& X=-v_1\gamma t+\frac{\left(v_1^2\gamma+v_2^2\right)x}{v^2}+\frac{v_1v_2\left(\gamma-1\right)y}{v^2}~,~~~  \\
&& Y=-v_2\gamma t+\frac{v_1v_2\left(\gamma-1\right)x}{v^2}+\frac{\left(v_1^2+v_2^2\gamma\right)y}{v^2}~,  \\
&& Z=z~.
\end{eqnarray}

\section{Equatorial geodesic equations of motion}  \label{sect3}
According to Eqs.~\eqref{g00} - \eqref{gij}, the weak-field equations of motion of test particles traveling in the equatorial plane $\left(z=\partial/\partial z=0\right)$ of the moving Schwarzschild source whose velocity is also constrained to the equatorial plane are obtained in the form
\begin{equation}
0=\ddot{t}+\frac{M\left(A_2\dot{t}^2-A_3\dot{x}^2-2A_4\dot{t}\dot{x}\right)}{A_1 R}+\mathcal{O}\left(M^2\right)~, \hspace*{6pt}   \vspace*{3pt}    \label{1PM-ddott}
\end{equation}
\begin{eqnarray}
&&0=\ddot{x}+\frac{M\left(A_5\dot{t}^2+A_6\dot{x}^2-2A_7\dot{t}\dot{x}\right)}{A_1 R}+\mathcal{O}\left(M^2\right)~,        \label{1PM-ddotx}  \\
&&0=\ddot{y}+\frac{M\left(A_8\dot{t}^2+A_9\dot{x}^2+2A_{10}\dot{t}\dot{x}\right)}{A_1 R}+\mathcal{O}\left(M^2\right)~,~~~~   \label{1PM-ddoty}
\end{eqnarray}
where $R=\sqrt{X^2+Y^2}$, a dot denotes the derivative with respect to the affine parameter $\xi$ which describes the trajectory of the particle~\cite{WS2004,We1972} and is assumed to take the dimension of length~\cite{WS2004}, $\dot{y}$ has been regarded as a first-order quantity~\cite{HL2016b}, and the coefficients are as follows:
\begin{widetext}
\begin{eqnarray}
&&A_1=\left[v^2t^2+\left(1-v_2^2\right)x^2+\left(1-v_1^2\right)y^2+2v_1v_2xy-2\left(v_1x+v_2y\right)t\right]^2~,  \\
\nn&&A_2=\left[R\!+\!\gamma \left(v^2t\!-\!v_1x\!-\!v_2y\right)\right]\!\big{\{}\!\left(v^2t\!-\!v_1x\!-\!v_2y\right)R\!+\!\gamma\big{[}v^4t^2\!
+\!\left(v_1^2\!-\!2v_2^2\gamma^{-2}\right)x^2\!+\!\left(v_2^2\!-\!2v_1^2\gamma^{-2}\right)y^2     \\
&&\hspace*{25pt}+\,2v_1v_2\left(3-2v^2\right)xy-2v^2\left(v_1x+v_2y\right)t\big{]}\big{\}}~,     \\
\nn&& A_3=v^2\left(2\!-\!v_1^2\!+\!2v_2^2\right)t^3\!+v_1\left(1\!-\!v_2^2\right)x^3\!-v_2\left(4\!-\!v_1^2\right)y^3\!-v_1\left(4\!-\!3v_1^2\right)t^2x\!-v_2\left(4\!+\!3v_2^2\!+\!5v^2\right)t^2y  ~~~~~~~~~
\end{eqnarray}
\begin{eqnarray}
\nn&&\hspace*{25pt}-\,2\left(v_1^2-v_2^2\gamma^{-2}\right)\!\gamma t^2R+\!\left(2+v_2^2v^2-3v^2\right)x^2t+\!v_2\left(v_1^2-\gamma^{-2}\right)\!x^2y+\!\left(2+10v_2^2+v_1^2v^2\right)y^2t \\
\nn&&\hspace*{25pt}-\,2\left(1\!-\!v_2^2\right)^2\!\gamma x^2R-\!v_1\!\left(2\!+\!v_2^2\!+\!v^2\right)y^2x
-\!2\left(v_1^2v_2^2\!-\!\gamma^{-2}\right)\!\gamma y^2R-\!2\left(1\!-\!v_2^2\right)\!\left(t\!-\!v_1x\!-\!v_2y\right)R^2     \\
&&\hspace*{25pt}+\,2v_1v_2\left(5-v^2\right)txy+4v_1\left(1-v_2^2\right)\gamma txR+4v_2\left(v_1^2-\gamma^{-2}\right)\gamma tyR-4v_1v_2\left(1-v_2^2\right)\gamma xyR~,  \\
\nn&& A_4=\left[R+\gamma\left(v^2t-v_1x-v_2y\right)\right]\!\big{\{}\!\left[v_1t-\left(1-v_2^2\right)x-v_1v_2y\right]R+\gamma\big{[}v_1v^2t^2-\left(2v_1^2+3v_2^2\gamma^{-2}\right)tx   \\
&&\hspace*{25pt}+\,v_1v_2\left(1-3v^2\right)ty+v_1\left(1-v_2^2\right)x^2+v_2\left(3-2v_2^2-v^2\right)xy+v_1\left(v_2^2-2\gamma^{-2}\right)y^2\big{]}\big{\}}~,  \\
\nn&& A_5=-3v_1v^2t^3+\!\left(3\!-\!v_2^2\!-\!2v_2^4\right)x^3+\!v_1v_2\left(1\!+\!2v_1^2\right)y^3+\!\left(6v_1^2\!-\!v_2^2v^2\!+\!3v^2\right)t^2x+\!v_1v_2\left(6\!+\!v^2\right)t^2y   \\
\nn&&\hspace*{25pt}+\,2v_1v^4\gamma t^2R\!-\!v_1\!\left(9\!-\!2v_2^2\right)x^2t\!+\!v_1v_2\!\left(1\!+\!6v_2^2\right)x^2y
\!+\!2v_1\!\left(v_1^2\!-\!v_2^2\gamma^{-2}\right)\!\gamma x^2R\!-\!v_1\!\left(3\!+\!2v_2^2\right)y^2t     \\
\nn&&\hspace*{25pt}+\left(3-\!v_2^2-\!6v_1^2v_2^2\right)y^2x+\!2v_1\left(v_2^2-\!v_1^2\gamma^{-2}\right)\gamma y^2R+\!2\left[v_1t-\!\left(1-\!v_2^2\right)x-\!v_1v_2y\right]R^2-\!2v_2txy     \\
&&\hspace*{25pt}\times\left(3+v_1^2-v_2^2\right)-4v_1^2v^2\gamma txR-4v_1v_2v^2\gamma tyR+4v_1^2v_2\left(2-v^2\right)\gamma xyR~,     \\
\nn&& A_6=\left[v_1R+\gamma\left(v_1t-\left(1-v_2^2\right)x-v_1v_2y\right)\right]\!\big{\{}v_1\!\!\left[v_1t-\!\left(1-v_2^2\right)\!x-v_1v_2y\right]\!R+\!\gamma\big{[}\!\left(v_1^2-2v_2^2\gamma^{-2}\right)\!t^2     \\
&&\hspace*{25pt}\,-2v_1\left(1\!-\!v_2^2\right)tx\!-\!2v_2\left(v_1^2\!-\!2\gamma^{-2}\right)ty\!+\!\left(1\!-\!v_2^2\right)^2x^2
\!+\!2v_1v_2\left(1\!-\!v_2^2\right)xy\!+\!\left(v_1^2v_2^2\!-\!2\gamma^{-2}\right)y^2\big{]}\big{\}}~, \hspace*{0.6cm}   \\
\nn&& A_7=\left[v_1R+\gamma\left(v_1t-\left(1-v_2^2\right)x-v_1v_2y\right)\right]\!\big{\{}v_1\left(v^2t-v_1x-v_2y\right)R+\gamma\big{[}v_1v^2t^2+v_1\left(1-v_2^2\right)x^2  \\
&&\hspace*{25pt}+\,v_1\left(v_2^2-2\gamma^{-2}\right)y^2-\left(2v_1^2+3v_2^2\gamma^{-2}\right)tx+v_1v_2\left(1-3v^2\right)ty+v_2\left(3-2v_2^2-v^2\right)xy\big{]}\big{\}}~,  \\
\nn&& A_8=-3v_2v^2t^3+v_1v_2\left(1+2v_2^2\right)x^3+\left(1+v_1^2\!-\!2v_1^4\right)y^3+v_1v_2\left(6+v^2\right)t^2x+\left(7v_2^2+v_1^2\gamma^{-2}\right)\!t^2y   \\
\nn&&\hspace*{25pt}+\,2v_2v^4\gamma t^2R-v_2\left(3\!+\!2v_1^2\right)x^2t\!+\!\left[3v_2^2\left(1\!-\!2v_1^2\right)\!
+\!\gamma^{-2}\right]x^2y\!+\!2v_2\left(v_1^2\!-\!v_2^2\gamma^{-2}\right)\gamma x^2R\!-\!v_2y^2t  \\
\nn&&\hspace*{25pt}\times\left(5-2v_1^2\right)-\!3v_1v_2\left(1\!-\!2v_1^2\right)y^2x+\!2v_2\left(v_2^2\!-\!v_1^2\gamma^{-2}\right)\gamma y^2R\!+\!2v_2\left(t\!-\!v_1x\!-\!v_2y\right)R^2\!-\!2v_1txy    \\
&&\hspace*{25pt}\times\left(2v_2^2\!+\!\gamma^{-2}\right)-4v_1v_2v^2\gamma txR-4v_2^2v^2\gamma tyR+4v_1v_2^2\left(2-v^2\right)\gamma xyR~,    \\
\nn&& A_{9}=-v_2\left(3v_2^2+v^2\right)t^3\!-\!v_1v_2\left(1-v_2^2\right)x^3\!+\!\left(1-v_1^2\right)\left(2+v_2^2+v^2\right)y^3\!-\!\left(v_1^2\!-\!10v_2^2\!-\!v_1^4\!-\!2v_2^4\right)t^2y       \\
\nn&&\hspace*{25pt}+\,v_1v_2\!\!\left(2\!+\!3v_2^2\!-\!v^2\right)\!t^2x\!+\!2v_2\!\!\left(v_1^2\!-\!v_2^2\gamma^{-2}\right)\!\gamma t^2R\!+\!v_2\!\!\left(v_1^2\!-\!\gamma^{-2}\right)\!x^2t\!-\!\!\left[\!\left(v_1^2\!-\!2\gamma^{-2}\right)\!v_2^2\!+\!\gamma^{-2}\right]\!x^2y         \\
\nn&&\hspace*{25pt}+\,2v_2\left(1\!-\!v_2^2\right)^2\gamma x^2R\!-\!v_2\left(8\!-\!5v_1^2\!+\!4v_2^2\right)y^2t\!+\!v_1v_2\left(v_2^2\!+\!3v^2\right)y^2x\!+2v_2\left(v_1^2v_2^2-\gamma^{-2}\right)\gamma y^2R          \\
\nn&&\hspace*{25pt}-\,2v_1\left(4v_2^2-\gamma^{-2}\right)txy\!-\!4v_1v_2\left(1-v_2^2\right)\gamma txR\!-\!4v_2^2\left(v_1^2-\gamma^{-2}\right)\gamma tyR\!+\!4v_1v_2^2\left(1-v_2^2\right)\gamma xyR      \\
&&\hspace*{25pt}+\,2v_2\left(1-v_2^2\right)\left(t-v_1x-v_2y\right)R^2~,              \\
\nn&& A_{10}=v_1v_2v^2t^3\!-v_2\left(1\!-\!v_2^2\right)x^3-3v_1\left(1\!-\!v_1^2\right)y^3-v_2\left(3v_1^2\!-\!2v_2^2\right)t^2x-5v_1v_2^2t^2y-2v_1v_2v^2\gamma t^2R     \\
\nn&&\hspace*{25pt}+\,v_1v_2\left(3\!-\!4v_2^2\right)x^2t+v_1v_2^2x^2y-2v_1v_2\left(1\!-\!v_2^2\right)\gamma x^2R\!+\!v_1v_2\left(7\!-\!4v_1^2\right)y^2t\!+\!v_2\left(2\!-\!5v_1^2\right)y^2x      \\
\nn&&\hspace*{25pt}-\,2v_1v_2\!\left(v_2^2\!-\!\gamma^{-2}\right)\!\gamma y^2R
\!-\!2v_2\!\left[v_1t\!-\!\left(1\!-\!v_2^2\right)x\!-\!v_1v_2y\right]\!R^2\!-\!4v_2^2\!\left(1\!-\!2v_1^2\right)txy\!+\!4v_1v_2^2v^2\gamma tyR    \hspace*{0.7cm}      \\
&&\hspace*{25pt}+\,4v_2v^2\left(1-v_2^2\right)\gamma txR-4v_2^2\left(1-v_2^2\right)\gamma xyR~.
\end{eqnarray}

\end{widetext}
\begin{figure*}
\centering
\begin{minipage}[b]{\textwidth}
\includegraphics[width=16cm]{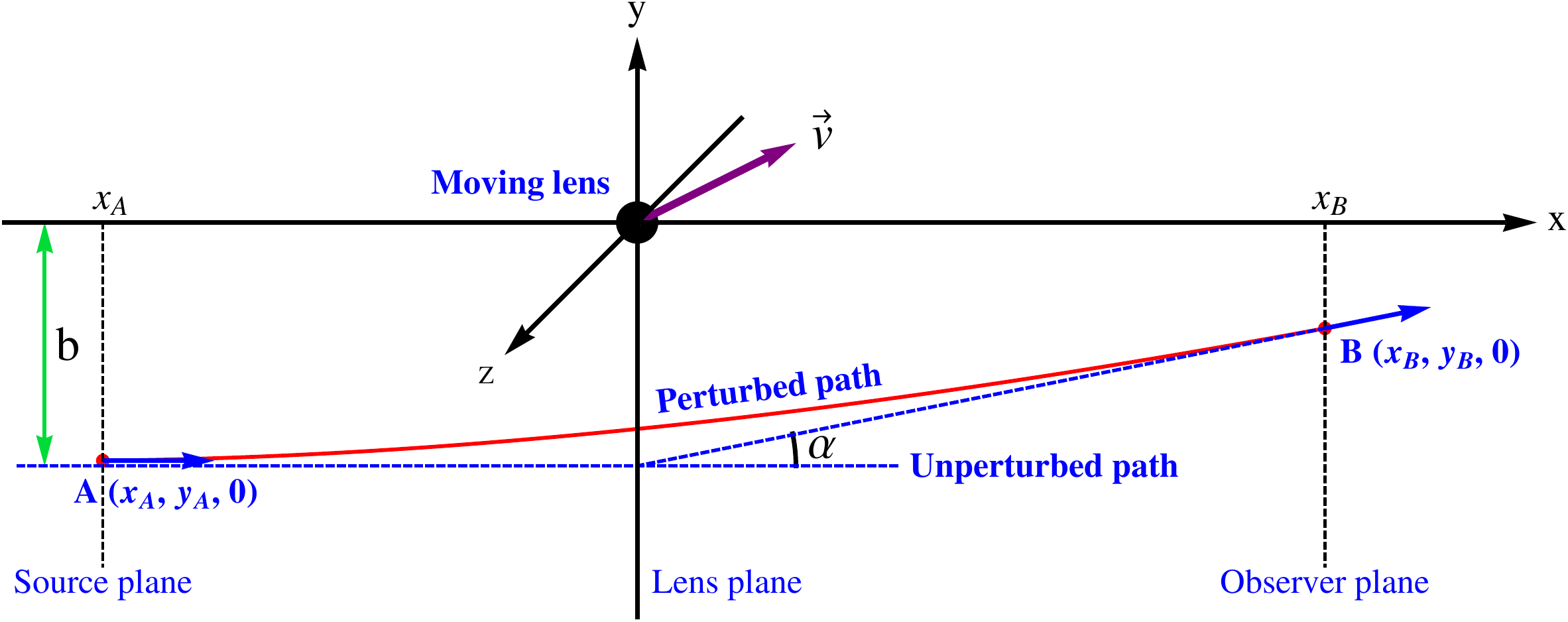}
\caption{Geometrical diagram for the propagation of a relativistic particle in the equatorial plane of a moving Schwarzschild black hole with an arbitrary two-dimensional constant velocity
$\bm{v}=v_1\bm{e}_1+v_2\bm{e}_2$. For guaranteeing the validity of the weak-field approximation, we only consider the scenario where $v_2\geq0$. Both the particle source $A$ and the observer $B$
are assumed to lie in the asymptotically flat region of the spacetime, and their spatial coordinates in the background's rest frame are denoted by $(x_A,~y_A,~0)$ and $(x_B,~y_B,~0)$, respectively,
where $y_A\approx-b$, $x_A\rightarrow -\infty$, and $x_B\rightarrow+\infty$, with $b$ being the impact parameter. In the comoving frame, their coordinates are denoted respectively as $(X_A,~Y_A,0)$ and
$(X_B,~Y_B,0)$. The red line represents the propagation trajectory of the particle traveling from $x\rightarrow-\infty$ with a relativistic initial velocity
$\bm{w}|_{x\rightarrow -\infty}\left(\approx\bm{w}|_{x\rightarrow x_A}\right)=w\bm{e}_1$ ($v_1<w$~\cite{WS2004}), while the dashed horizontal line stands for its flight trajectory when the lens is absent. For the convenience of our discussion, a particle with $0.05\lesssim w$ is roughly regarded to be relativistic. The deflection angle of the particle is denoted by $\alpha$, and the gravitational deflection effect is exaggerated greatly to distinguish its perturbed path from the unperturbed one. }   \label{Figure1}
\end{minipage}
\end{figure*}

\section{Weak deflection of relativistic particles by the moving Schwarzschild source} \label{sect4}
In this section, we focus on computing the gravitational deflection up to the 1PM order of relativistic particles including photons by the moving Schwarzschild lens, and then discuss the influence of the transversal motion of the lens on the leading-order Schwarzschild deflection of the particles. Figure~\ref{Figure1} presents the corresponding geometrical diagram for a relativistic test particle traveling from the source $A$ to the receiver $B$ in the field of the moving Schwarzschild lens.

We first consider the calculation of the leading-order equatorial deflection angle of a relativistic particle via an iterative technique developed in~\cite{WS2004,HL2017b}. The gravitational bending angle is defined by the difference of the propagation directions of the test particle before and after passing the lens, which can be written within the 1PM approximation as~\cite{HL2017b}
\begin{widetext}
\begin{eqnarray}
\alpha\equiv\arctan\!\left.\frac{\dot{y}}{\dot{x}}\right|^{\xi\rightarrow +\infty}_{\xi\rightarrow -\infty}
=\arctan\!\left.\frac{\dot{y}}{\dot{x}}\right|^{x\rightarrow +\infty}_{x\rightarrow -\infty}
=\left.\frac{\dot{y}}{\dot{x}}\right|^B_A+\mathcal{O}\left(M^2\right)~.~~~     \label{angle-0}
\end{eqnarray}
\end{widetext}
Up to the 0PM order, Eqs.~\eqref{1PM-ddott} - \eqref{1PM-ddoty} yield
\begin{eqnarray}
&& \dot{t}=\frac{1}{w}+\mathcal{O}\left(M\right)~,  \label{0PM-dott}   \\
&& \dot{x}=1+\mathcal{O}\left(M\right)~,  \label{0PM-dotx}   \\
&& \dot{y}=0+\mathcal{O}\left(M\right)~,  \label{0PM-doty}
\end{eqnarray}
where the boundary conditions $\dot{t}|_{\xi\rightarrow -\infty}=\dot{t}|_{x\rightarrow -\infty}=\frac{1}{w}$~, $\dot{x}|_{\xi\rightarrow -\infty}=\dot{x}|_{x\rightarrow -\infty}=1$~, and $\dot{y}|_{\xi\rightarrow -\infty}=\dot{y}|_{x\rightarrow -\infty}=0$ have been used. Additionally, the combination of Eqs.~\eqref{0PM-dott} and \eqref{0PM-dotx} gives the 0PM expression of $t$
\begin{eqnarray}
&&t=\frac{x}{w}+\mathcal{O}\left(M\right)~,     \label{0PM-t}
\end{eqnarray}
where a zero value of the integration constant has been adopted. Moreover, Eqs.~\eqref{0PM-dotx} - \eqref{0PM-doty}, together with the boundary condition $y|_{\xi\rightarrow -\infty}=y|_{x\rightarrow -\infty}=-b$, lead to the following 0PM forms of $y$ and the parameter transformation
\begin{eqnarray}
&&y=-b+\mathcal{O}\left(M\right)~,     \label{0PM-y} \\
&&dx=\left[1+\mathcal{O}\left(M\right)\right]d\xi~.     \label{0PM-PT}
\end{eqnarray}

We next substitute Eqs.~\eqref{0PM-dott} - \eqref{0PM-PT} into the integration of Eq.~\eqref{1PM-ddoty} over $\xi$, and obtain up to the 1PM order
\begin{widetext}
\begin{eqnarray}
\dot{y}=\frac{\gamma \left[\left(B_2x+B_3\right)B_1+B_4x^3+B_5x^2+B_6x+B_7\right]M}{B_1^3}+\,\mathcal{O}\left(M^2\right)~,    \label{1PM-doty}
\end{eqnarray}
with
\begin{eqnarray}
&&B_1=\sqrt{\left[v_2^2+\left(v_1-\left(1-v_2\right)w\right)\left(v_1-\left(1+v_2\right)w\right)\right]x^2+2v_2w\left(1-v_1w\right)bx
+\left(1-v_1^2\right)w^2b^2}~,  \\
&&B_2=2v_2\left[\left(v_1-w\right)^2+v_2^2\left(1-w^2\right)\right]~,      \\
&&B_3=2v_2^2w\left(1-v_1w\right)b~,      \\
&&B_4=\frac{\left[1+v^2-4v_1w+\left(1+v_1^2-v_2^2\right)w^2\right]
\left[v_2^2+\left(v_1-\left(1-v_2\right)w\right)\left(v_1-\left(1+v_2\right)w\right)\right]}{wb}~, \\
&&B_5=2v_2\left(1-v_1w\right)\left[1+3v^2-8v_1w+\left(3+v_1^2-3v_2^2\right)w^2\right]~,  \\
&&B_6=w\left[1+8v_2^2+\left(1-2v_1^2\right)v^2-2v_1\left(3-3v_1^2+7v_2^2\right)w+\left(2-v^2-v_2^2-v_1^4+9v_1^2v_2^2\right)w^2\right]b~,    \\
&&B_7=4v_2w^2\left(1-v_1w\right)\left(1-v_1^2\right)b^2~.
\end{eqnarray}
\end{widetext}
Finally, by substituting Eqs.~\eqref{0PM-dotx} and \eqref{1PM-doty} into Eq.~\eqref{angle-0} and considering the conditions $x_A\ll -b$ and $x_B\gg b$, the leading-order deflection angle of relativistic particles (including massive and massless particles) traveling from the source to the observer in the equatorial plane of a moving Schwarzschild black hole which has an two-dimensional equatorial velocity can be obtained in a compact form
\begin{eqnarray}
\alpha\left(v_1,\,v_2,\,w\right)=2\,N\left(1+\frac{1}{w^2}\right)\frac{M}{b}+\mathcal{O}\left(M^2\right)~,~~~   \label{Angle}
\end{eqnarray}
where the kinematical correctional coefficient $N$ contains the joint effect of the radial and transversal lens motions and is given by
\begin{eqnarray}
N\!\left(v_1,\,v_2,\,w\right)\!=\!\frac{w\gamma\!\left[\left(1\!+\!v^2\right)\!\left(1\!+\!w^2\right)\!-\!4v_1w\!-\!2v_2^2w^2\right]}{\left(1\!+\!w^2\right)\!\sqrt{\left(1-v_2^2\right)w^2+v^2-2v_1w}} ,~~~~~~~    \label{N}
\end{eqnarray}
with $0.05\lesssim w\leq 1$, $v_1<w$~\cite{WS2004}, and $0\leq v_2<\sqrt{1-v_1^2}$.

It can be seen that for the scenario of no transversal motion of the lens $(v_2=0)$, the kinematical coefficient becomes $N\left(v_1,~0,~w\right)=\left(1-\frac{v_1}{w}\right)^{-1}\left(1+v_1^2-\frac{4v_1 w}{1+w^2}\right)\gamma_1$, where $\gamma_1=\left(1-v_1^2\right)^{-\scriptstyle\frac{1}{2}}$, and Eq.~\eqref{Angle} reduces to the first-order gravitational deflection angle of relativistic particles caused by a radially moving Schwarzschild source~\cite{HL2017b}
{\small
\begin{eqnarray}
\alpha\left(v_1,~0,~w\right)=\frac{2\,\gamma_1\!\left[\left(1\!+\!v_1^2\right)\!\left(1\!+\!\frac{1}{w^2}\right)
\!-\!\frac{4v_1}{w}\right]\!M}{\left(1-\frac{v_1}{w}\right)b}\!+\!\mathcal{O}\left(M^2\right)\,. ~~~~~~~  \label{Angle-1}
\end{eqnarray}}
On the other hand, for the case of no radial lens motion $(v_1=0)$, the Lorentz factor becomes $\gamma_2=\left(1-v_2^2\right)^{-\scriptstyle\frac{1}{2}}$, and the leading-order equatorial deflection of relativistic particles due to a moving Schwarzschild lens with a transversal velocity is thus achieved as
{\small\begin{eqnarray}
\alpha\left(0,~v_2,~w\right)=\frac{2\gamma_2\!\left[1+w^2+v_2^2\left(1-w^2\right)\right]\!M}{w\sqrt{w^2+v_2^2\left(1-w^2\right)}\,b}+\mathcal{O}\left(M^2\right)~,~~~~~~~   \label{Angle-2}
\end{eqnarray}}
from which we see that no effect of the transversal lens motion appears up to the first order in velocity~\cite{PB1993,WS2004}. However, when the second-order terms of the velocity $v_2$ are considered, the transversal motion effect of the lens will be present and Eq.~\eqref{Angle-2} can be expanded as
\begin{eqnarray}
&&\nn\alpha\left(0,~v_2,~w\right)=2\left(1+\frac{1}{w^2}\right)\!\left[1+\frac{v_2^2\left(3w^2-1\right)}{2w^2\left(1+w^2\right)}\right]\!\frac{M}{b}  \\
&&\hspace*{2.23cm}+\,\mathcal{O}\left(M^2,~v_2^3\right)~.   \label{Angle-3}
\end{eqnarray}

Moreover, the leading-order gravitational deflection of light $(w=1)$ by a transversally moving Schwarzschild lens can be obtained according to Eq.~\eqref{Angle-2}
\begin{eqnarray}
\alpha\left(0,~v_2,~1\right)=\frac{4\gamma_2M}{b}+\mathcal{O}\left(M^2\right)~.   \label{Angle-4}
\end{eqnarray}
It is also interesting to find that in the limits $v_1\rightarrow 0$ and $v_2\rightarrow 0$, Eq.~\eqref{Angle} is consistent with the result of the weak-field Schwarzschild deflection of relativistic particles in the literature~\cite{AR2002,AR2003,HL2016a,CG2018,PJ2019,JBGA2019}, since the kinematical coefficient $N\left(0,~0,~w\right)$ recovers $1$ directly.

As to the kinematical coefficient $N$ given in Eq.~\eqref{N}, it should be mentioned that its deviation from 1 is mainly caused by the motion of the lens in the background's rest frame $\left(t,~x,~y,~z\right)$. Since a boosted observer who is at rest in the lens' comoving frame $(T,~X,~Y,~Z)$ moves relative to the static observer in $\left(t,~x,~y,~z\right)$, this deviation may also be partially interpreted as a result of the change in the propagation path of a test particle due to the aberration effect originated from the motion of the boosted observer relative to the static one.

\section{Discussion of velocity effects} \label{sect5}
Now we consider further the influences of the transversal and radial motions of the lens on the weak-field deflection of relativistic massive particles and light in the context of general relativity. Based on Eq.~\eqref{Angle} and the notations adopted in~\cite{HL2016a}, the joint transversal and radial velocity-induced effect on the first-order Schwarzschild deflection of relativistic particles is defined as
\begin{eqnarray}
\Delta\alpha\left(v_1,~v_2,~w\right)\equiv\alpha\left(v_1,\,v_2,\,w\right)-\alpha\left(0,\,0,\,w\right)~.~~~~   \label{DeltaAngle}
\end{eqnarray}
It should be mentioned that when the lens' radial velocity vanishes ($v_1=0$), Eq.~\eqref{DeltaAngle} denotes a pure transversal velocity effect on the Schwarzschild deflection
\begin{widetext}
\begin{eqnarray}
\Delta\alpha\left(0,~v_2,~w\right)=2\left\{\frac{w\left[\left(1+v_2^2\right)\left(1+w^2\right)-2v_2^2w^2\right]}
{\left(1+w^2\right)\sqrt{1-v_2^2}\sqrt{\left(1-v_2^2\right)w^2+v_2^2}}-1\right\}\left(1+\frac{1}{w^2}\right)\frac{M}{b}+\mathcal{O}\left(M^2\right)~.~~~~   \label{DeltaAngle-2}
\end{eqnarray}
\end{widetext}
If the transversal motion of the lens is omitted, Eq.~\eqref{DeltaAngle} then reduces to the radial velocity effect $\Delta\alpha\left(v_1,~0,~w\right)$ on the deflection angle of relativistic particles including photons, which has been discussed in~\cite{HL2016a} in detail.

\begin{figure*}
\centering
\begin{minipage}[b]{\textwidth}
\includegraphics[width=16cm]{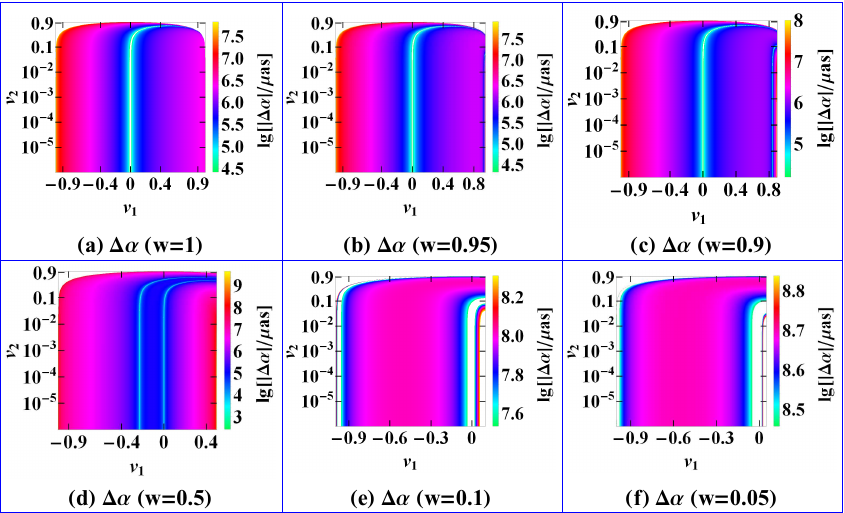}
\caption{$\Delta\alpha\left(v_1,~v_2,~w\right)$ plotted in color-indexed form for various $v_1$ and $v_2$, with different initial velocities of the relativistic particle. Here and thereafter, we present the
cases with high values of $v_1$ and $v_2$ mainly for illustration, since the possibility that a massive black hole moves at a relativistic velocity is very small or even non-existent. Moreover, we focus
on the absolute values of the joint velocity effect. }   \label{Figure2}
\end{minipage}
\end{figure*}

\begin{figure*}
\centering
\begin{minipage}[b]{\textwidth}
\includegraphics[width=16cm]{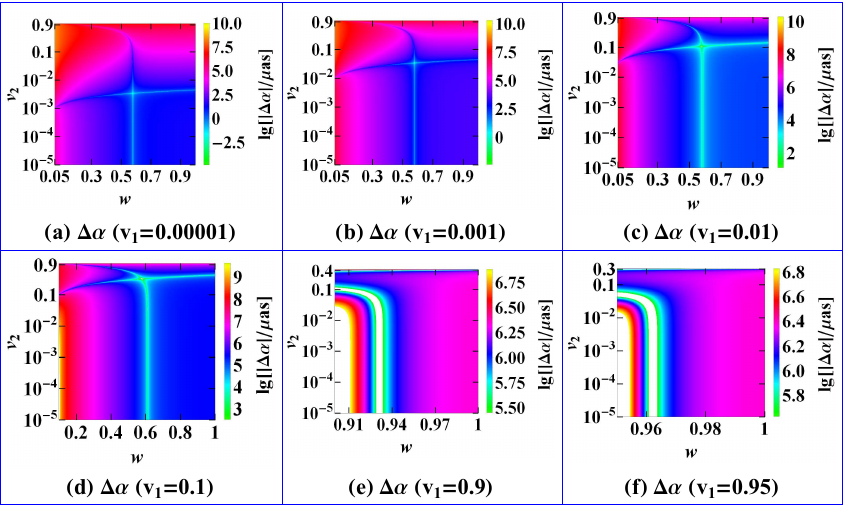}
\caption{Color-indexed joint transversal and radial velocity effect shown as a function of $w$ and $v_2$ $(0\leq v_2<\sqrt{1-v_1^2}\hspace*{0.5pt})$ for various radial lens velocities.  }   \label{Figure3}
\end{minipage}
\end{figure*}

\begin{figure*}
\centering
\begin{minipage}[b]{\textwidth}
\includegraphics[width=16cm]{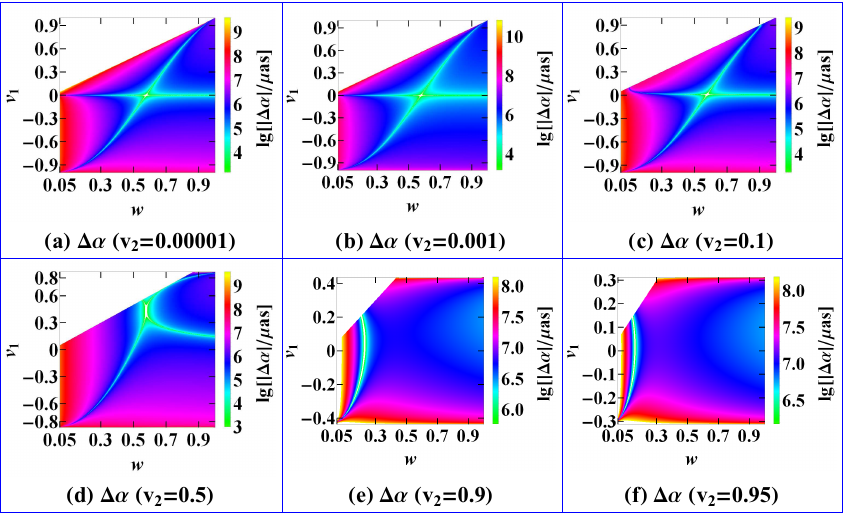}
\caption{Color-indexed joint transversal and radial velocity effect plotted as a function of $w$ and $v_1$ for different transversal lens velocities, with $-\sqrt{1-v_2^2}<v_1<\min\,\{w,~\sqrt{1-v_2^2}\hspace*{0.5pt}\}$. }   \label{Figure4}
\end{minipage}
\end{figure*}

For estimating the order of magnitude of the velocity effects, the moving lens is assumed to have a rest mass $M\!=\!4.2\times10^{6}M_{\odot}$~\cite{BG2016,Parsa2017} of the galactic supermassive black hole (i.e., Sgr A$^\ast$) as an example, with $M_{\odot}$ denoting the rest mass of the Sun. Additionally, we follow the idea of~\cite{IRDMA2008,HZFMWPL2020} and approximately regard the impact parameter $b$ as the Einstein radius $R_E=5.74\times10^{-5}\,$kpc of Sgr A$^\ast$ for the special case of $d_L=d_{LS}$, where $d_L$ $(=8.2\,$kpc~\cite{BG2016}$)$ and $d_{LS}$ stand for the observer-lens and lens-source angular diameter distances, respectively. Figure~\ref{Figure2} shows the joint transversal and radial velocity-induced effect $\Delta\alpha\left(v_1,~v_2,~w\right)$ as a function of the velocity components $v_1$ and $v_2$ in color-indexed form for various fixed $w$ (i.e., for different relativistic particles). With respect to the results given in Fig.~\ref{Figure2}, two aspects should be mentioned. First, Fig.~\ref{Figure2} (a) explicitly displays that the absolute values of $\Delta\alpha\left(v_1,~v_2,~1\right)$ for photons ($w=1$) acting as test particles are at least four orders of magnitude larger than the planned angular accuracy of the SKA~\cite{BBGKW2015,LXLWBLYHL2022} in almost all relativistic-motion cases of the lens. Although the lens moves at a nonrelativistic velocity $v$ $(|v_1|\ll 1,~|v_2|\ll 1)$, $|\Delta\alpha\left(v_1,~v_2,~1\right)|$ can exceed $1\mu$as evidently under many scenarios. For instance, the value of $|\Delta\alpha\left(v_1,~v_2,~1\right)|$ is about $28.9\mu$as in the nonrelativistic case $v_1=v_2=1.0\times10^{-5}$, and it reaches $30.3\mu$as when $v_1=-1.0\times10^{-5}$ and $v_2=0.001$ are assumed. Thus, the possibility to detect the joint transversal and radial velocity effect on the leading-order Schwarzschild deflection of light is large with current angular resolution under proper astronomical scenarios, whether the lens' motion is relativistic or not. Second, the residual subfigures of Fig.~\ref{Figure2} (namely, Fig.~\ref{Figure2} (b) - (f)) indicate that the possibility to measure the joint velocity effect for the case of a relativistic massive particle being the test particle is tiny or even non-existent for all cases of the lens' motion nowadays, because of the low angular resolution ($\lesssim 1^{\circ}$) of current particle and multi-messenger detectors~\cite{SW2009,Aab2014,Albert2020}. For example, the value of $|\Delta\alpha\left(v_1,~v_2,~w\right)|$ can reach about $2.06\,$as for a high-energy cosmic-ray particle with a relativistic initial velocity $w=0.1$, when $v_2=0.01$ and $v_1=-0.001$ are given. This value is still much smaller than current angular resolution of the particle detectors, although it is six and eight orders of magnitude larger than the aimed astrometric accuracies of the SKA and the mentioned NEAT mission~\cite{Malbet2012,Malbet2014}, respectively. It is only when the radial component $v_1$ of the lens velocity gets close enough to $w$ and the transversal velocity $v_2$ approaches a tiny value that $|\Delta\alpha\left(v_1,~v_2,~w\right)|$ may be larger than the angular accuracy of current particle detectors. For instance, $|\Delta\alpha\left(v_1,~v_2,~w\right)|$ approaches $3.94^{\circ} (>1^{\circ})$, if we assume $w=0.5$, $v_1=0.4999$, and $v_2=0.0001$. However, considering the rapid progress made in techniques and instruments of astronomical measurements, we can expect that the joint transversal and radial velocity effect on the leading-order Schwarzschild deflection of massive particles is also likely to be observed via future high-accuracy multi-messenger detectors. Finally, for the reader's convenience, we also exhibit $\Delta\alpha\left(v_1,~v_2,~w\right)$ as a function of $w$ and $v_2$ in Fig.~\ref{Figure3} for different radial lens velocities and as a function of $w$ and $v_1$ for various transversal velocities of the lens in Fig.~\ref{Figure4} in color-indexed form.

\section{Summary} \label{sect6}
In summary, a new compact expression for the leading-order gravitational deflection of relativistic massive particles and light propagating in the time-dependent field of a moving Schwarzschild black hole with a two-dimensional equatorial velocity has been achieved iteratively, within the framework of the first post-Minkowskian approximation. Since the kinematical correctional coefficient of the resulting bending angle includes not only the radial motion effect of the lens but also the effect of the transversal lens motion, our result can be regarded as a natural extension of the ones in the literature. We have also discussed the joint transversal and radial velocity effect on the leading-order Schwarzschild deflection of the relativistic particles including photons, and the order of magnitude of this kinematical effect and the possibility of detecting it have been estimated. In most cases, it is found that the possibility to detect the joint velocity effect on the leading-order deflection of light signals with current astrometric resolution is large. In contrast to it, the detection of the joint velocity effect on the first-order Schwarzschild deflection of relativistic massive particles is beyond the limited capability of current particle detectors, and depends on future high-accuracy multi-messenger detectors.

\section*{Acknowledgments}
This work was supported in part by the National Natural Science Foundation of China (Grant Nos. 12205139 and 12475057).

\end{document}